**Living Liquid Crystals**

Shuang Zhou[a,1], Andrey Sokolov[b,1], Oleg D Lavrentovich[a, 2] , Igor S Aranson[b,c,2]

[a]*Liquid Crystal Institute and Chemical Physics Interdisciplinary Program, Kent State University, Kent, OH 44242, USA*

[b] *Materials Science Division, Argonne National Laboratory, Argonne, IL60439*

[c] *Engineering Sciences and Applied Mathematics, Northwestern University, 2145 Sheridan Road, Evanston, IL 60202, USA*

**Key words:** active nematic, lyotropic chromonic liquid crystals, bacteria, flagella motion, flow, instability, correlation length,

The authors declare no conflict of interest.

[1] S.Z. and A.S. contributed equally to this work.

[2] To whom correspondence should be addressed. E-mail: aronson@anl.gov, or olavrent@kent.edu

**Abstract**

Collective motion of self-propelled organisms or synthetic particles often termed "active fluid" has attracted enormous attention in broad scientific community because of it fundamentally non-equilibrium nature. Energy input and interactions among the moving units and the medium lead to complex dynamics. Here we introduce a new class of active matter – living liquid crystals (LLCs) that combine living swimming bacteria with a lyotropic liquid crystal. The physical properties of LLCs can be controlled by the amount of oxygen available to bacteria, by concentration of ingredients, or by temperature. Our studies reveal a wealth of new intriguing dynamic phenomena, caused by the coupling between the activity-triggered flow and long-range orientational order of the medium. Among these are (a) non-linear trajectories of bacterial motion guided by non-uniform director, (b) local melting of the liquid crystal caused by the bacteria-produced shear flows, (c) activity-triggered transition from a non-flowing uniform state into a flowing one-dimensional periodic pattern and its evolution into a turbulent array of topological defects; (d) birefringence-enabled visualization of microflow generated by the nanometers-thick bacterial flagella. Unlike their isotropic counterpart, the LLCs show collective dynamic effects at very low volume fraction of bacteria, on the order of 0.2%. Our work suggests an unorthodox design concept to control and manipulate the dynamic behavior of soft active matter and opens the door for potential biosensing and biomedical applications.

**Significance Statement**

We propose a new class of active matter, the living liquid crystal (LLC), representing motile rod-shaped bacteria placed in water based non-toxic liquid crystal. Long-range orientational order of the liquid crystal and the swimming activity of bacteria demonstrate a strong coupling that dramatically alters individual and collective bacterial dynamics. For example, swimming bacteria perturb the orientational order of the liquid crystal or even cause its local melting, making the flagella motion optically visible. Second, self-organized textures emerge from the initial uniform LLC alignment with a characteristic length controlled by a balance between bacteria activity and anisotropic viscoelastisity of liquid crystal. Third, the local liquid crystal orientation controls the direction of motion of bacteria. LLC can leads to valuable biosensoring and biomedical applications.

Active matter has recently emerged as an important physical model of living systems that can be described by the methods of nonequilibrium statistical mechanics and hydrodynamics [1-3]. The active matter is driven by the internal sources of energy, associated with the self-propelled particles such as bacteria or synthetic swimmers. The interaction of these active particles among themselves and with the medium produces a rich variety of dynamic effects and patterns. Most of the studies deal with active particles embedded into a Newtonian isotropic fluid. In this case the interactions among particles are caused by long-range hydrodynamic and short-range excluded volume effects [4-13]. In this work, we conceive a new general class of active fluids, termed living liquid crystals (LLCs). The suspending medium is a non-toxic liquid crystal that supports the activity of self-propelled particles, namely, bacteria. At the very same time, the medium imposes long-range anisotropic interactions onto bacteria, thanks to the intrinsic orientational order that persists even when the bacteria are not active. The importance of the new system is two-fold. Firstly, the bacterial activity modifies the orientational order of the system, by producing well-defined and reproducible patterns with or without topological defects. Secondly, the orientational order of the suspending medium reveals new facets of bacterial behavior, allowing one to control trajectories of individual bacteria and to visualize rotation of flagella through birefringence of the host. The LLCs represent a new example of a biomechanical system, capable of controlled transduction of stored energy into a systematic movement, which is of critical importance in a variety of application, from bioinspired micromachines to self-assembled microrobots (14,15). The study of bacterial motion in LCs and non-Newtonian fluids takes us a step closer to realizing *in vitro* environments that more closely resemble conditions *in vivo* (16,17).

**Results**

The concept of LLC is enabled by the recent progress in water-soluble non-toxic chromonic LCs (16-18) and growing expertise in control and manipulation of bacterial suspensions in confined geometries (10-14). Our studies have shown that *living bacteria can be transferred to the LC media, and yield highly nontrivial interactions with the molecular ordering of the LC*. The experiments are conducted with common swimming bacteria (*Bacillus subtilis)* in lyotropic chromonic, see Materials and Methods. We examined the simplest nematic phase of the LLC. In the absence of activity, the LLC is a standard nematic characterized by the long-range orientational order described by a unit director $\hat{\mathbf{n}}$ with the property $\hat{\mathbf{n}} = -\hat{\mathbf{n}}$. Its ground state is a uniform director field, $\hat{\mathbf{n}}$ =const. When activity is turned on, the LLC exhibits the onset of large-scale undulations of a nematic director with a characteristic length $\xi$ (see below) determined by the balance between bacteria activity and anisotropic viscoelasticity of the lyotropic chromonic. The nematic phase of the LLC depends on both the concentration of the mesogenic material and temperature; at low concentrations and/or high temperatures, the material melts into an isotropic fluid. The bacterial-free chromonic LC and the LLC show approximately the same phase diagram in the concentration-temperature coordinates. In an experimental cell, the direction of $\hat{\mathbf{n}}$ is preselected by surface anchoring, namely, by a rubbed layer of polyimide coated onto the glass substrates (19), see Materials and Methods; this direction is along the *x*-axis, so that the unperturbed director is $\hat{\mathbf{n}}_0 = (1,0,0)$.

A typical bacterium has a cylindrical body of length 5-7 μm and diameter 0.7 μm. There are also about 20 helicoidal 10 μm long flagella filaments attached to the bacterial body. In an active (motile) bacterium, the filaments bundle on one end of the bacterium and rotate, thus powering unidirectional "head-forward" motion. The motility of aerobic bacteria (such as *Bacillus subtilis*) is controlled by the amount of dissolved oxygen (10,13,14). When placed in the uniformly aligned LC, the bacteria show a number of intriguing dynamic phenomena that can be attributed to the coupling of the LC structure to an individual bacterium and to the collective effects. Since the individual behavior is pertinent for the understanding of emerging collective motion, we first discuss the dynamics of individual bacteria in relatively thin (of thickness $h = 5$ μm) LLC flat glass cells, Fig 1.

Because of the rod-like shape of the bacterium, and because of the strong orienting effect of the surface anchoring in a thin LC cell, individual bacteria tend to swim along director $\hat{\mathbf{n}}_0$, in agreement with earlier observations (17, 20). The most striking feature observed in our experiments is that the bacterial flagella, having a diameter of only about 24 nm, produces director perturbations on a scales of micrometers, Fig. 1A,B,C. Birefringence of the LLC makes these perturbations clearly seen under the polarizing microscope, Fig. 1A. The texture in Fig. 1A observed with de-crossed polarizers, reveals periodic dark and bright elongated spots tilted with respect to $\hat{\mathbf{n}}_0$ and caused by the broken chiral symmetry of the director distortions (see SI). The map of optical retardance in Fig. 1B, obtained with a PolScope (see Materials and Methods) (21), demonstrates that the effective birefringence near the bacterium is reduced as compared to the uniform director surrounding.

The pattern of alternating bright and dark spots in the bacterium wake propagates with a wavelength about $d = 2$ μm. At a fixed distance from the body of bacterium, the bright and dark regions alternate with the frequency of about 16 Hz, Fig. 1C. The wavelength is determined by the pitch of helical flagella, and the frequency by the flagella rotation rate. We constructed a space-time diagram by stacking a total of 240 consecutive cross-sections along the bacterium's axis in a co-moving reference frame from each image, Fig. 1D. From the space-time diagram we clearly see propagation of the flagella wave (parallel white lines between 0 and 15 μm) and counter-rotation of the bacterial body (dark regions at -5 μm) with a rate of 2.5 Hz. The ratio of flagella rotation to the counter-rotation of the body is about 7:1, similar to that known for *Bacillus subtilis* under normal conditions (160 Hz flagella rotation and 20 Hz body counter-rotation (22)).

Individual behavior of bacteria and its coupling to the orientational order becomes especially interesting as the temperature is increased and the LLC approaches a biphasic region, in which the isotropic and nematic phases coexist. The isotropic regions appear as characteristic "negative tactoids" elongated along the overall director of the surrounding nematic (23,24). The isotropic tactoids, seen as dark islands in Fig. 1E, distort the director around them and change the trajectories of the bacteria. As shown in Fig.1E, far away from the tactoid, a bacterium is swimming along a straight line set by the uniform director. In the vicinity of tactoid, the trajectory deviates from the straight line and follows the local distorted director. After a collision with the isotropic-nematic interface, the bacterium follows the curved interface, and finally escapes at the cusp of the tactoid.

Even more strikingly, the bacteria themselves can create isotropic tactoids in their wake if the LLC temperature is close to the biphasic region, Fig. 1F. The LLC acts as miniature "Wilson chamber" in which the isotropic droplets decorate the path of swimming bacteria. The feature underscores the complexity of interplay between velocity fields and the state of orientational order that can involve a number of different mechanisms, such as existence of a nucleation barrier, non-uniform distribution of components between the nematic and isotropic phase, etc. Nucleation of the isotropic phase in Fig. 1F implies that the bacterial flows reduce the local degree of orientational order, most probably through disintegration of the chromonic aggregates (18). The fact that bacteria can follow the nematic-isotropic interface and the overall director in the LLC offers numerous design concepts of reconfigurable microfluidic devices for the control and manipulation of bacteria. The desired trajectories of bacterial transport can be engineered by patterned surface anchoring and by local dynamic heating (for example, with focused laser beams).

Now we discuss collective behavior of LLC emerging at higher concentrations of bacteria. We discovered that a long-range nematic alignment of the LLC is affected by the flow created by the swimming bacteria. The coupling of the orientational order and hydrodynamic flow yields nontrivial dynamic patterns of the director and bacterial orientations, see Figs. 2 and 3, and SI.

The first example, Fig. 2, demonstrates the existence of a characteristic spatial length scale $\xi$ in LLCs that sets this non-equilibrium system apart from standard equilibrium LCs. The LLC is confined between two glass plates that fix $\hat{\mathbf{n}}_0 = (1, 0, 0)$ along the rubbing direction. In the samples with inactive bacteria the steady state is uniform, $\hat{\mathbf{n}} = \hat{\mathbf{n}}_0 = \text{const}$, and the immobilized bacteria are aligned along the same direction, Fig. 2A,B. In chambers with active bacteria, supported by the influx of oxygen through the air-LLC interface (on the left hand side in Fig. 2E), the uniform state becomes unstable and develops a stripe pattern with periodic bend-like deviations of $\hat{\mathbf{n}}$ from $\hat{\mathbf{n}}_0$, Fig. 2C-E. The swimming bacteria are aligned along the local director $\hat{\mathbf{n}}$, Fig. 2D. Since the oxygen permeates the LLC from the open side of the channel, its concentration is highest at the air-LC interface; accordingly, the stripes appear first near the air-LLC interface. The period $\xi$ of stripes increases with the increase of the distance from the air-LLC interface as the amount of oxygen and the bacterial activity decrease. Figure 2H shows that $\xi$ increases when the concentration $c$ of bacteria and the chamber height $h$ decrease. The data in Fig. 2H are collected for different samples in which the velocity of bacteria was similar ($8\pm3$ μm/s), as established by a particle-tracking velocimetry; the concentration is normalized by the concentration $c_0 \approx 8 \times 10^8$ cells/cm$^3$ of the stationary growth phase.

As time evolves, in the regions with the highest bacterial activity, near the open edge, the stripe pattern becomes unstable against nucleation of pairs of $\pm\frac{1}{2}$ disclinations, Fig. 2F, G. Remarkably, the pattern-forming instabilities occurring here have no direct analog for bacterial suspensions in Newtonian fluids or for bacteria-free pure LCs. The concentration of bacteria in our experiments (close to 0.2% of volume fraction) is about 1/10 of that needed for the onset of collective swimming in Newtonian liquids (about $10^{10}$ cells/cm$^3$ (10)).

Figure 3 illustrates the profound effect of bacterial activity on spatio-temporal patterns in a sessile drop of LCCs (10,25) in which there is no preferred director orientation in the plane of film. Bacterial activity generates persistently rearranging bacterial and director patterns with ±½ disclinations that nucleate and annihilate in pairs, similarly to the recent experiments on active microtubule bundles (4), Fig.3A. The characteristic spatial scale of the pattern, determined as an average distance between the disclination cores, is in the range of 150-200 μm, of the same order of magnitude as $\xi$ in the stripe pattern in strongly anchored sample. The fluid flow typically encircles disclination pairs, Figs. 3B-D.

**Discussion**

Emergence of a characteristic length scale $\xi$ in LCCs, either as a period of the stipe pattern in Fig. 2 or as a characteristic separation of disclinations in Fig. 3, is caused by the balance of director-mediated elasticity and bacteria-generated flows, Fig. 2I. Since no net force is applied to a self-propelled object, a swimming bacterium represents a moving negative hydrodynamic force dipole of the strength $U_0$ ("pusher"), as it produces two outward fluid streams coaxial with the bacterial body (26). The strength of the dipole (of a dimension of torque or energy) is of the order 1 pN μm (27). In the approximation of nearly parallel orientation of the bacterium and the local director $\hat{\mathbf{n}}$, the bacteria-induced streams, Fig.2I, impose a reorienting torque $\sim\alpha(h)cU_0\theta$, where $\alpha(h)$ is a dimensionless factor that describes the flow strength and depends on the cell thickness $h$; $c$ is the concentration of bacteria, and $\theta$ is the angle between local orientation of bacteria and $\hat{\mathbf{n}}$. Similar torques caused by shear flow are well known in the physics of liquid crystals (24), see SI and Fig.S3. Mass conservation yields an estimate $\alpha=\alpha_0 l/h$, where constant $\alpha_0\approx O(1)$, and $l$ is the length of a bacterium. It implies that the channel's thickness reduction increases the flow because the bacteria pump the same amount of liquid. The local bacterium-induced reorienting hydrodynamic torque is opposed by the restoring elastic torque $\sim K\frac{\partial^2\theta}{\partial x^2}$; $K$ is an average Frank elastic constant of the LC. In the case of a very thin layer confined between two plates with strong surface anchoring, the strongest elastic torque $K_2\frac{\partial^2\theta}{\partial z^2}$ will be associated with the twist deformation along the vertical $z$-axis. However, since the elastic constant for twist is an order of magnitude smaller than the splay and bend constants (28), the restoring torque in relatively thick (20 and 50 $\mu$m) samples is caused mainly by in-plane distortions. By balancing the elastic and bacterial torques, $\frac{\partial^2\theta}{\partial x^2} = \frac{\theta}{\xi^2}$, one defines a "bacterial coherence length" $\xi = \sqrt{\frac{Kh}{\alpha_0 lcU_0}}$ (somewhat similar arguments for the characteristic length of bending instability in active nematics were suggested in [2]). This expression fits the experimental data on the periodicity of stripe patterns for different concentrations of bacteria $c$ and thicknesses $h$ remarkably well with a choice of $\alpha_0 \approx 1$, see Fig. 2H. A simple phenomenological theory shows that competition between bacteria-generated hydrodynamic torque and restoring elastic torque yields nontrivial pattern-forming instability (see SI and Fig. S4).

The experiments in Fig.2 and 3 clarifies the rich sequence of instabilities by which the activity of bacteria transforms the initial non-flowing homogeneous uniform steady state in Fig.2A,B into the state of self-sustained active fluid turbulence, Fig.3. The first step is through appearance of the periodic director bend with a characteristic length scale $\xi$, Fig.2C-E. The period becomes shorter as the activity increases, Fig.2H. Further escalation of the activity causes a qualitative transformation, namely, nucleation of defect pairs, Fig.2F,G. Note that the axis connecting the cores of ½ and –½ disclinations in the pair and the local director $\hat{\mathbf{n}}$ along this axis are perpendicular the original director $\hat{\mathbf{n}}_0$, Fig.2G. Once the system can overcomes the stabilizing action of surface anchoring, as in Fig.3, the dynamic array of moving defects forms a globally isotropic state in which the local director $\hat{\mathbf{n}}$ is defined only locally (somewhat similar dynamic behavior was observed in simulations of "active nematic" in Refs. [29,30]). The final state, Fig.3, is an example of "active fluid" turbulence at vanishing Reynolds number [10,25], which is in our case on the order of $10^{-5}$.

In conclusion, LLC demonstrates a wealth of new phenomena not observed for either suspension of bacteria in a Newtonian fluid or in passive ordered fluid. The concept of a characteristic length $\xi$, contrasting the elastic response of orientationally ordered medium and the activity of microswimmers, may also be useful for understanding hierarchy of spatial scales in other active matter systems (6-11). Our studies were focused on the simplest nematic LLC. However, more complex LLCs can be explored as well, for example, smectics and cholesteric LCs with controlled chirality. Exploration of LLCs may have intriguing applications in various fields. Our biomechanical system may provide the basis for devices with new functionalities, including specific responses to chemical agents, toxins, or photons. Swimming bacteria can also serve as autonomous "microprobes" for the properties of LCs. Unlike passive microprobes (31), swimming bacteria introduce local perturbations of the LC molecular order both in terms of the director and the degree of order, and thus provide unique information on the mesostructure of the material. In turn, LC medium provides valuable optically accessible information on the intricate submicrometer structure of bacteria-generated microflow that deserve further investigation.

## Materials and Methods

**<u>Bacteria.</u>** Experiments were conducted with the strain 1085 of *Bacillus subtilis*, rod-shaped bacterium 5 μm long and 0.7 μm in diameter. The bacteria were initially grown on LB agar plates, then transferred to a Terrific Broth liquid medium (Sigma T5574) and grown in shaking incubator at temperature 35C for 10-12 hours. To increase resistance to oxygen starvation the bacteria are grown in sealed vials under microaerobic conditions (10). We monitored the concentration of bacteria by the measurement of the optical scattering of the growth media. The bacteria at the end of their exponential growth stage were extracted and washed. The growth medium was separated from the bacteria and removed as completely as possible by centrifugation.

**<u>LLC preparation</u>.** Chromonic lyotropic LC material disodium cromoglycate (DSCG) purchased from Spectrum Chemicals, 98% purity, was dissolved in Terrific Broth at 16 wt%. This solution was added to the concentrated bacteria obtained as described above. The resulting LLC was mixed by stirring with a clean toothpick and then in a vortex mixer at 3000 rpm. The LLC mixture was injected (by pressure gradient) into a flat cell made from two square glass plates (10×10 mm) separated by spacers. Two interior surfaces were pretreated with polyimide SE7511 and rubbed with velvet cloth to provide a uniform planar alignment of the LLC. The cell was sealed with high vacuum grease (Dow Corning) to prevent water evaporation. The temperature shift of the LC phase diagrams during several hours of the experiments is less than 1˚C and thus does not affect the data presented. The observations were started immediately after the sealed cells were placed in a heating stage (Linkam PE94) at 25˚C. For the used lyotropic LC, the average value of the splay and bend elastic constants is $K$=10-12 pN (28); the average viscosity $\eta$~10 kg/(m s) as determined in an experiment with director relaxation in cells with magnetically induced director distortions (Frederiks effect ) (24, 28).

**Videomicroscopy.** An inverted microscope Olympus IX71 with a motorized stage, mounted on a piezoelectric insolation platform Herzan TS-150 and Prosilica GX 1660 camera (resolution of 1600x1200) were used to record motion of individual bacteria in thin cells. Images were acquired with the frame rate up to 100 frames/sec, at 60x magnification, oil-immersion objective, in cross-polarized light. Color camera with the resolution 1280x1024 and the frame rate 10 frames/second and 2x-20x magnifications were used to acquire large-scale patterns of collective motion. The acquired images were processed in Matlab.

**PolScope microscopy**. The LLC textures were examined by a polarizing microscope (Nikon E600) equipped with Cambridge Research Incorporation (CRI) Abrio LC-PolScope package. The LC PolScope uses a monochromatic illumination at 546 nm and maps optical retardance $\Gamma(x, y)$ in the range (0-273) nm and orientation of the slow axis (21). For tangentially anchored LLC, $\Gamma = |n_e - n_0|h$ , where $n_e$ and $n_o$ are the extraordinary and ordinary refractive indices, $h$ is the cell thickness. For DSCG, optical birefringence is negative, $n_e - n_0 \approx -0.02$ (18). The slow axis is thus perpendicular to the optic axis and to $\hat{\mathbf{n}}$. The PolScope was set up to map the local orientation $\hat{\mathbf{n}}(x, y)$, Fig. S2b.

**Acknowledgements**

The research of AS and ISA was supported by the US DOE, Office of Basic Energy Sciences, Division of Materials Science and Engineering, under the Contract No. DE AC02-06CH11357. ODL and SZ were supported by the NSF grant DMR 1104850.

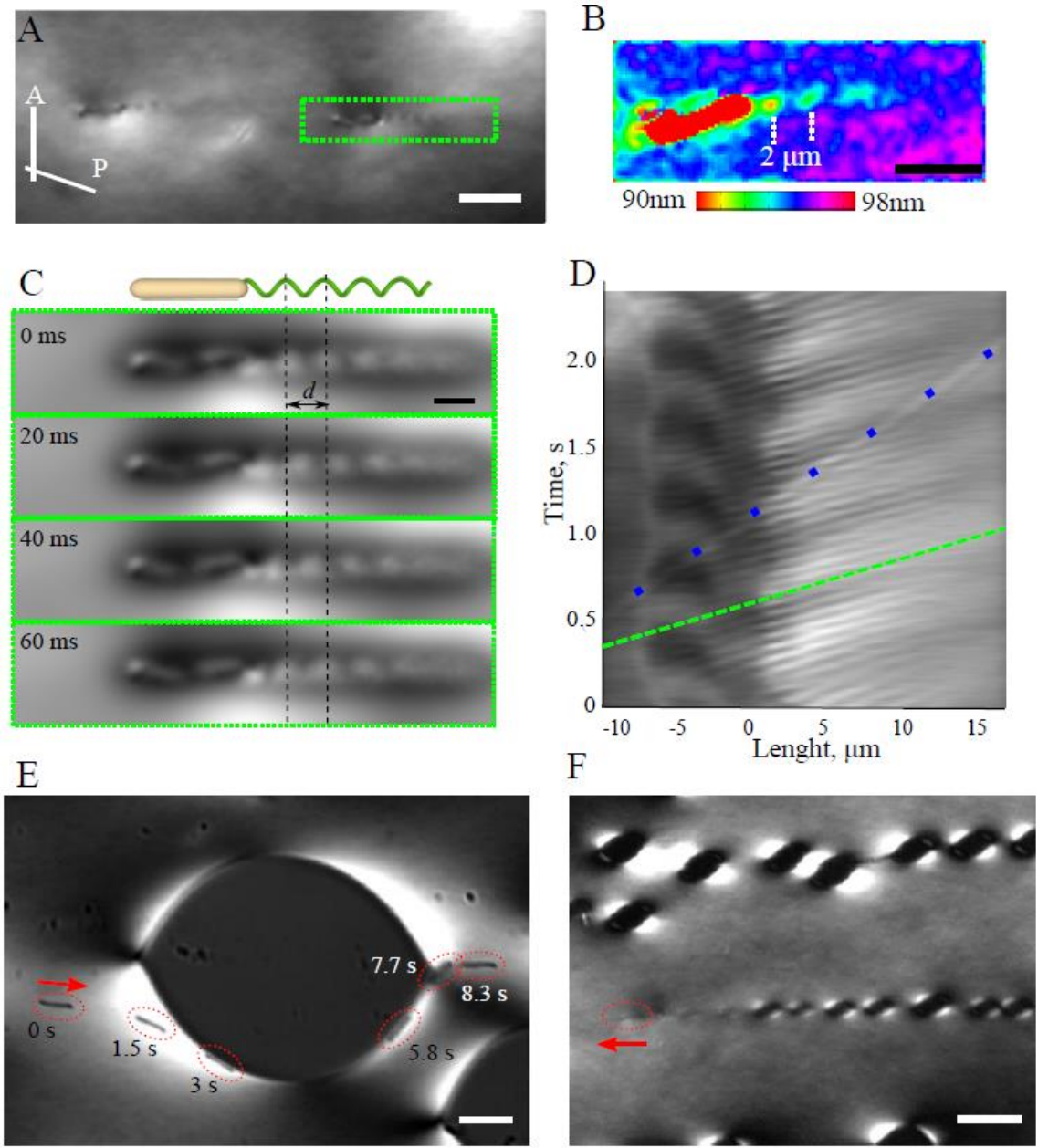

**Figure 1. Distortion of the nematic director detected by optical microscopy.** (A)

Snapshot of swimming bacteria observed under a microscope with slightly de-crossed polarizer (P) and analyzer (A). The bacterium shown in green box swims from the right to left.  (B) Optical retardance pattern around a swimming bacterium, see also SI Figures S1- S3. (C) Time evolution of the director waves created by rotating flagella in the co-moving reference frame..  (D) Space-time diagram for director waves extracted for the bacterium shown in panel (C). A total of 240 cross-sections were extracted from 2.4 sec video.  Dashed green line depicts phase velocity of the flagella wave. Dots mark an immobilized dust particle.  See also SI Movies 1 & 2. (E)  The trajectory of a single bacterium around a tactoid. (F) Trace of isotropic tactoids left by a bacterium at temperature about  0.5 ˚C below the nematic-biphasic transition point. Observations are made under a microscope with slightly de-crossed polarizer (P) and analyzer (A). Scale bar 5 μm (A,B), 2 μm (C), 10 μm (E), 20 μm (F). See SI Movies 3 & 4.

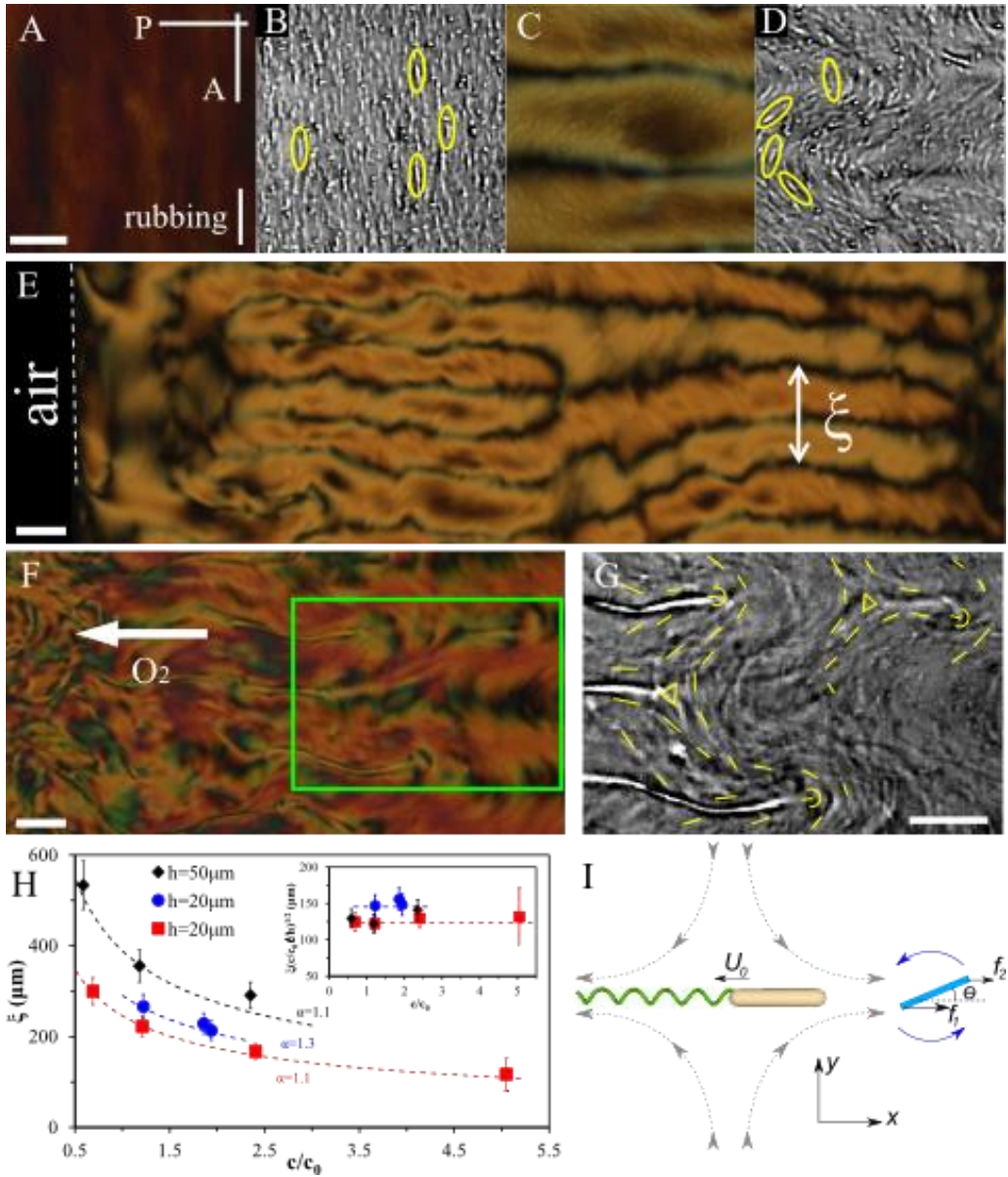

**Figure 2. Emergence of a characteristic length scale in LLCs.** (A,B) LLC with inactive bacteria is at its equilibrium state with the director and bacteria (highlighted by ellipses) aligned uniformly along the rubbing direction; (C,D) active bacteria produce periodically distorted director. (E) Proliferation of stripe pattern in the sample of thickness $h$=20 μm and for low concentration of bacteria, $c \approx 0.9 \times 10^9$ cells/cm$^3$. Oxygen

permeates from the left hand side. (F) LLC patterns in thicker sample ($h$=50 μm) and for higher concentration of bacteria, $c \approx 1.6 \times 10^9$ cells/cm$^3$. White arrow points toward a higher concentration of oxygen. (G) Zoomed area in panel (F) showing nucleating disclinations of strength +1/2 (semi-circles) and -1/2 (triangles). Bright dashes visualize bacterial orientation. (H) Dependence of characteristic period $\xi$ on $c$ and $h$; dashed lines depict fit to theoretical prediction $\xi = \sqrt{\dfrac{Kh}{\alpha_0 l c U_0}}$. Inset illustrates collapse of the data into a universal behavior that follows from the theoretical model. (I) Director realignment (shown as a rod) caused by the bacterium-generated flow (shown by dashed lines with arrows). See also SI Movies 5 & 6. Scale bar 50 μm (A-D), 100 μm (E-G).Error bars are ±10% SEM (standard error of the mean), except for ±30% SEM at c/c$_0$=5.05.

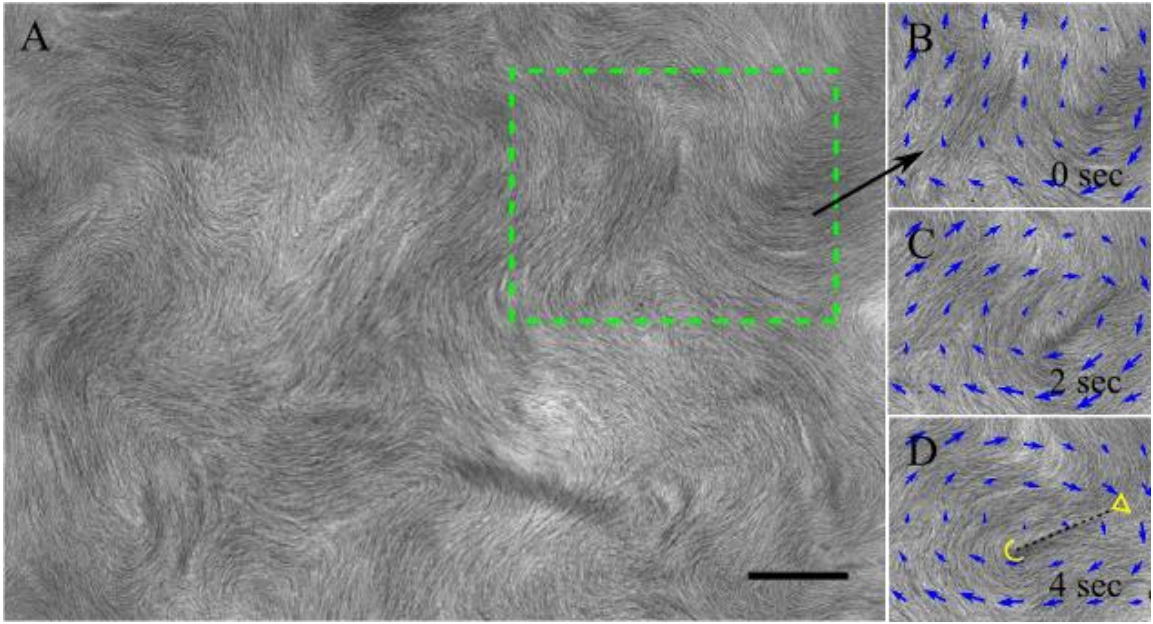

**Figure 3. LLC in sessile drop.** (A) Texture with multiple disclination pairs, green rectangle indicates the region shown in (B,C,D). Bacteria are aligned along the local nematic director, as revealed by the fine stripes. Scale bar 30 μm. No polarizers. See also SI Movie 7. (B,C,D) LLC texture with -1/2 and 1/2 disclinations and the pattern of local flow velocity (blue arrows) determined by particle-image velocimetry. The flow typically encircles the close pair of defects.

**Supporting Information**

**Living Liquid Crystals**

Shuang Zhou[1], Andrey Sokolov[2], Oleg D Lavrentovich[1], Igor S Aranson[2,3]

[1]*Liquid Crystal Institute and Chemical Physics Interdisciplinary Programme, Kent State University, Kent, OH 44242, USA*

[2]*Materials Science Division, Argonne National Laboratory, Argonne, IL60439*

[3]*Engineering Sciences and Applied Mathematics, Northwestern University, 2145 Sheridan Road, Evanston, IL 60202, USA*

## 1. Optics of director patterns in the wake of moving bacterium.

We numerically simulated the optical patterns observed under the polarizing microscope, using standard Berreman 4x4 matrix method [1]. We model the director field distorted by the flagella as following:

$$n_y(x,y,z) = \sin\varphi_0 \sin(-kx)\left(1-\beta|z-z_0|\right)\exp\left(-\frac{|y-y_0|}{\lambda}\right) \quad,$$

$$n_z(x,y,z) = -\sin\varphi_0 \cos(-kx)\left(1-\beta|z-z_0|\right)\exp\left(-\frac{|y-y_0|}{\lambda}\right),$$

$$n_x(x,y,z) = \sqrt{1-n_y^2-n_z^2} \;,$$

where $(x,y_0,z_0) = (x,r_0\cos(-kx),r_0\sin(-kx))$ defines the position of left handed helical flagella along $x$-axis and centered at (0,0) of y-z plane. By its rotation, director $\hat{\mathbf{n}}$ deviates from $\hat{\mathbf{n}}_0 = (1,0,0)$ by the angle $\varphi_0 = 10°$ at the flagella and linearly decay in $z$ direction with the rates $\beta = \left(z_0+\dfrac{h}{2}\right)^{-1}$ (for $z<z_0$) and $\left(\dfrac{h}{2}-z_0\right)^{-1}$ (for $z \geq z_0$), and exponentially decay in the $y$ direction with a characteristic decay length $\lambda = 1$ μm; $r_0 = 0.5$ μm is the rotation radius of flagella; $k = 2\pi P^{-1}$, $P$=2 μm is the helical pitch of the flagella. By using this model, we display in Fig S1 the textures for different analyzer orientations in the area corresponding to four periods of the director pattern along the $x$ direction and to 3 $\lambda$ along the $y$ direction.

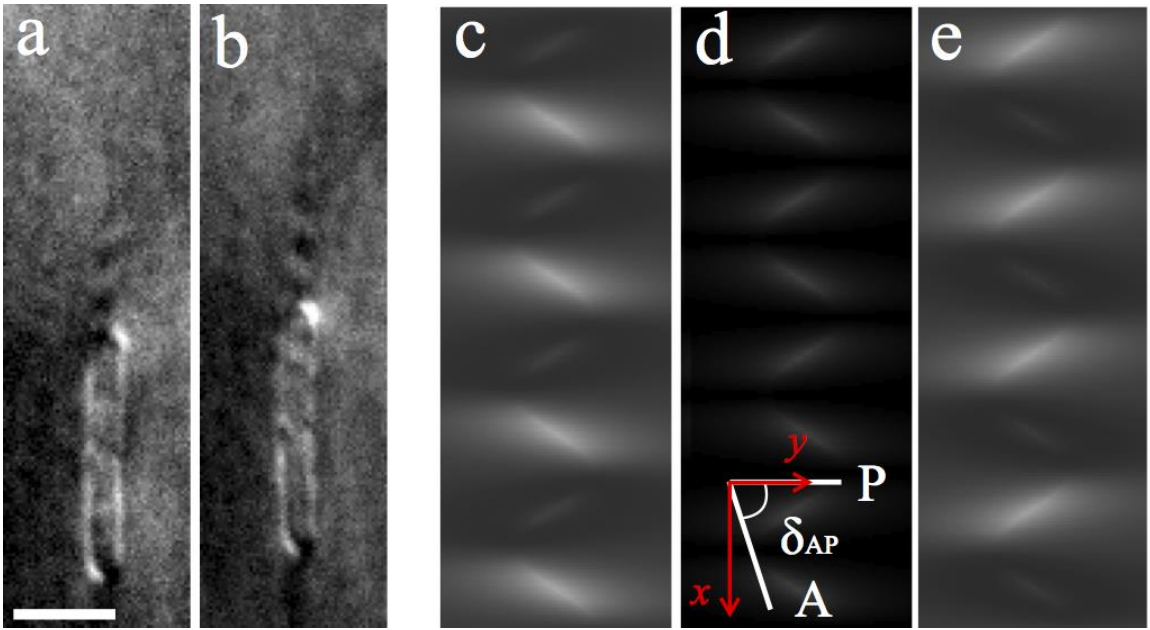

**Supplementary figure 1.** Light intensity pattern for a linearly polarized light passing through the twisted director configuration as described in the SI text. (a,b) polarizing microscopy textures of oblique black and white regions in the wake area of swimming bacteria. (c,d,e) Optical simulation of director textures deformed by the helicoidal flagella viwed between two linear polarizers making a different angle $\delta_{AP}$:=80°(c), 90°(d) and 100°(e). Scale bar 5µm.

## 2. Optics of director patterns distorted by flows produced by individual bacterium

Fig 1a shows that the regions under and above a moving bacterium have a different optical density. Fig. S2 shows a clear manifestation of this effect, for a bacterium that is pinned or does not swim (but otherwise active and has rotating flagella), apparently because it is close to the stage of division into two bacteria and has two sets of flagella. For this bacterium, the optical density pattern resembles a butterfly, Fig. S2a, in which the director distortions propagate over tens of micrometers. PolScope image of the director field indicates that the director forms a tilted pattern in the shape of the letter "X", Fig. S2b. The effect can be explained by the flow-induced reorientation of the director, schematically shown in Fig. S2c (see also Fig. 3i). Optical simulations based on the Berreman matrices show that these director distortions results in the butterfly pattern when the sample is observed between two polarizers, Fig. S2d,e,f. Analytical calculation with Jones matrix method shows that the field of light passing through a polarizer, a sample with retardance $\Gamma$ oriented at $\varphi$, and a analyzer at $\varphi_{AP}$ follows

$$\begin{bmatrix} E_a \\ 0 \end{bmatrix} = \frac{1}{\sqrt{2}} \begin{bmatrix} 1 & 0 \\ 0 & 0 \end{bmatrix} \begin{bmatrix} \cos\varphi_{AP} & \sin\varphi_{AP} \\ -\sin\varphi_{AP} & \cos\varphi_{AP} \end{bmatrix} \begin{bmatrix} \cos\varphi & -\sin\varphi \\ \sin\varphi & \cos\varphi \end{bmatrix} \begin{bmatrix} e^{-i\Gamma/2} & 0 \\ 0 & e^{i\Gamma/2} \end{bmatrix} \begin{bmatrix} \cos\varphi & \sin\varphi \\ -\sin\varphi & \cos\varphi \end{bmatrix} \begin{bmatrix} 1 \\ 0 \end{bmatrix}$$
,

thus intensity at the analyzer

$$\mathbf{I}_a = \frac{1}{2}\cos^2(\varphi_{AP} - \varphi)\cos^2\varphi + \sin^2(\varphi_{AP} - \varphi)\sin^2\varphi - \frac{1}{4}\sin 2(\varphi_{AP} - \varphi)\sin 2\varphi \cos\Gamma$$

For $\Gamma = -1.15$ ($h = 5\mu m$, $\Delta n = -0.02$, wavelength of the light 546nm) as in our samples, $\mathbf{I}_a$ has minimum values at $\varphi_{AP} \approx 90° + \frac{1}{2}\varphi$ when $\varphi < 25°$. This confirms the experimental and optical simulation texture of "butterfly" shaped director field, Fig. S2a,d,e,f For a moving bacterium, the effect is qualitatively similar, with the difference that the propulsion enhances the front two "wings" of the butterfly and weakens the rear two wings; as a result, the regions below and above the bacterium's head have different brightness, Fig. 1a,c.

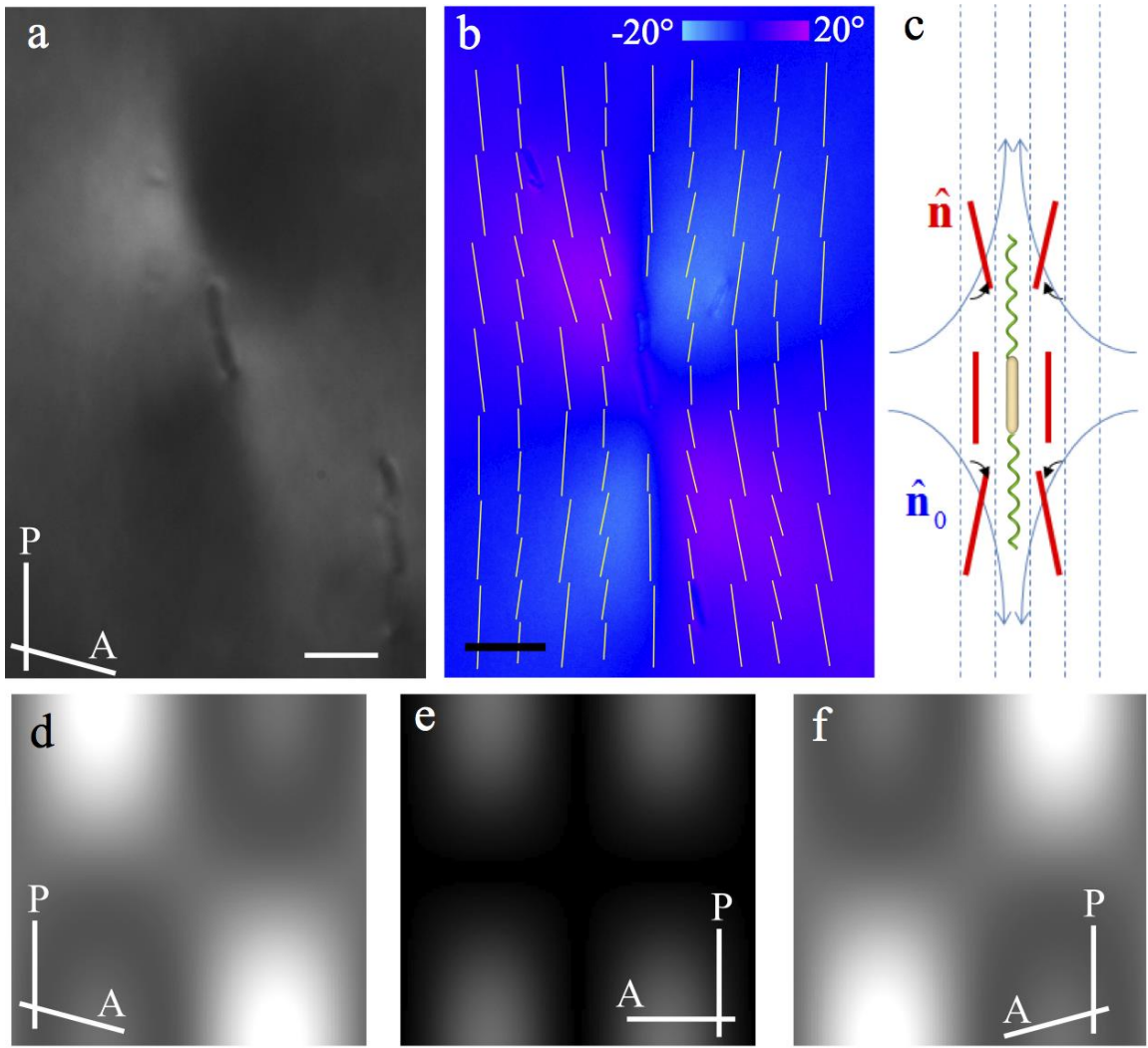

**Supplementary Fig 2:** Director distortions around an immobile bacterium. (a) Optical polarizing microscopy texture with de-crossed polarizers shows a "butterfly" pattern. (b) PolScope texture maps director pattern (yellow lines) resembling an "X" letter; color scale is proportional to the angle that the local director makes with the long axis of the image. (c) A schematic representation

showing how the director (red bar) deviates from $\hat{\mathbf{n}}_0 = (1,0,0)$ (blue dashed lines) due to the flow (blue arcs with arrows) induced by non-swimming two-tail bacterium. (d,e,f) Optical simulation of the director pattern in part (b) shows the butterfly pattern in the intensity map when the two polarizers are crossed at different angles (a,b). Scale bar 5 μm (a), 10 μm(b).

Besides the two coaxial streams, the bacterium also creates velocity fields associated with rotations of its body and its flagella (in opposite directions, Fig. S3a,b). These rotations cause director reorientations that lack the mirror symmetry with respect to the plane passing through the long axis of the bacterium, Fig. S3c; these deformations would further complicate the director pattern in the close vicinity of a moving bacterium; these effects will be explored in details in the future.

The director distortions would be created by a moving bacterium when the viscous torque overcomes the elastic one, which is equivalent to the requirement that the Ericksen number of the problem is larger than 1. For a rotating bacterium in a cell of thickness $h$, the Ericksen number [2] is $Er = \eta\, frh / K$, where $\eta$ is an effective viscosity, $f$ is the frequency of bacterium head rotation, $r$ is the bacterium radius. For typical values of parameters in our problem, $K = 10\,\text{pN}$, $\eta = 10\,\text{kg}/(\text{m}\cdot\text{s})$, $f = 2\,\text{Hz}$, $r = 0.4\,\mu\text{m}$, $h = 20\,\mu\text{m}$, this requirement is easily satisfied, as $Er$=16. As already discussed in the main text, the shear flows produced by the bacteria can also cause a decrease in the degree of orientational order, which corresponds to Deborah numbers close to 1 or larger.

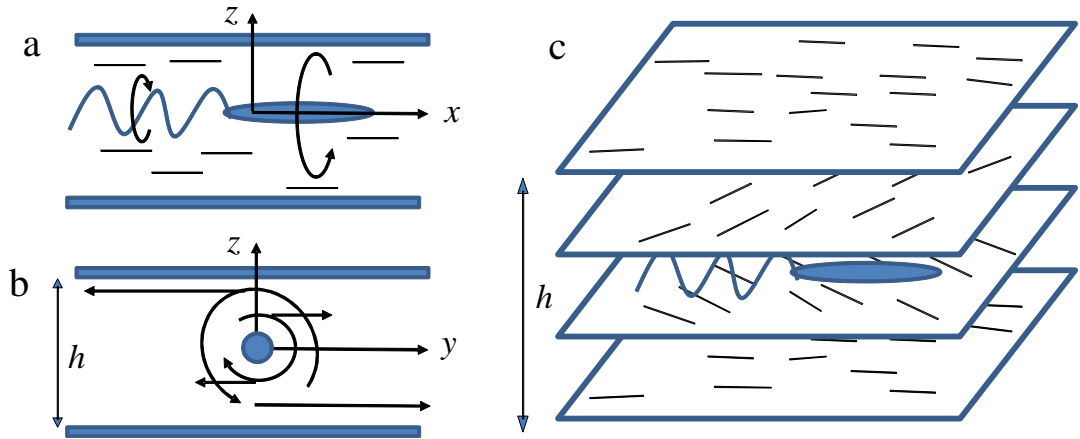

**Supplementary figure 3. Chiral symmetry breaking of the LLC director pattern caused by a rotating bacterium.** (a,b). Schematics of the flows generated by the rotating bacterium. (c) Scheme of the director twist along the vertical *z*-xis.

## 3. Pattern formation in the director dynamics: one-dimensional theory of LLC

Let $\theta$ and $\phi$ be corresponding angles between the nematic director and bacteria with respect to *x*-axis. All the quantities are averaged with respect to the thickness of the sample. For simplicity we focus on the one-dimensional situation (all quantitites depend only on *x*-coordinate). We postulate the following equations of motion following from the fact that the nematic tends to align bacteria, and the bacteria tend to rotate the nematic director from that direction:

$$\eta \partial_t \theta = \frac{U}{2} \sin(2(\phi - \theta)) + K \frac{\partial^2 \theta}{\partial x^2} - \varepsilon \sin(2\theta) \ (1)$$

$$\eta \partial_t \phi = -\frac{A}{2} \sin(2(\phi - \theta)) + \mu \frac{\partial^2 \phi}{\partial x^2} \ (2)$$

Here $U = cu_0 < 0$ is the averaged bacterial torque imposed on the director, $c$ is the concentration of bacteria, $u_0 < 0$ is the bacteria hydrodynamic dipole moment, $K$ is the liquid crystal elastic constant, $A$ is the rate of alignment of bacteria by the liquid crystal, and $\mu$ is the effective bacterial diffusion (we assume in the following that $\mu \gg K$ due to self-propulsion), and $\eta > 0$ is the rotational friction coefficient. The last term in Eq (1) describes an externally applied aligning (when $\varepsilon > 0$) or disorienting (when $\varepsilon < 0$) field that acts on the nematic. In a simplified one-dimensional model, this term can be qualitatively considered as the surface anchoring; for $\varepsilon = 0$, all director orientations in the plane of the sample are degenerate.

Equations (1,2) display surprisingly rich dynamics. Consider first the case $\varepsilon = 0$ (no externally imposed aligning field). Examine the stability of a fully coaxial case, $\theta = \phi$. We seek the solution to linearized Eqs. (1,2) near $\theta = \phi$ in the form $\theta, \phi \sim \exp(ikx + \lambda t)$. Here $k, \lambda$ are the modulation wavenumber and instability growth rate, respectively. Simple stability analysis shows that the system exhibits a long-wavelength instability. The growth rate $\lambda$ for small $k$ is of the form

$$\lambda(k) = -\frac{AK + \mu U}{\eta(A + U)} k^2 + O(k^4) (3)$$

Since the hydrodynamic dipole $U<0$, then $\lambda>0$ if $A>-U$ and $|U|>K\,A/\mu$. This instability leads to a rather non-trivial pattern formation. Further we examine Eqs. (1,2) numerically, with and without surface anchoring. Equations show very rich spatio-temporal dynamics as shown in Supplementary Fig. S4. If the externally imposed alignment is strong, no pattern appears and the fully aligned state of nematic and bacteria $\phi=\theta=0$ is stable. Reduction of the strength $\varepsilon$ of this externally imposed alignment (or, equivalently, increase of the bacterial activity torque $U$) leads to the onset of stripe pattern as shown in space-time diagram, Supplementary Fig. S4a. This regular stripe pattern seems to be stable (or, at least, very long-lived) only very close to the onset of instability. Further reduction of $\varepsilon$ leads to a gradual instability of the stripes via side undulations and the onset of a spatio-temporal intermittency, Supplementary Fig. S4b. In this case we observe patches of regular stripes mediated by randomly-oriented domains, similar to that what we see in experiment for high concentration of bacteria. Finally, at $\varepsilon=0$, we have randomly moving and oscillating domains with no preferred orientation, Supplementary Fig. S4c. This case likely corresponds to the behavior in sessile drop where in-plane orientation is degenerate, Fig. S5A.

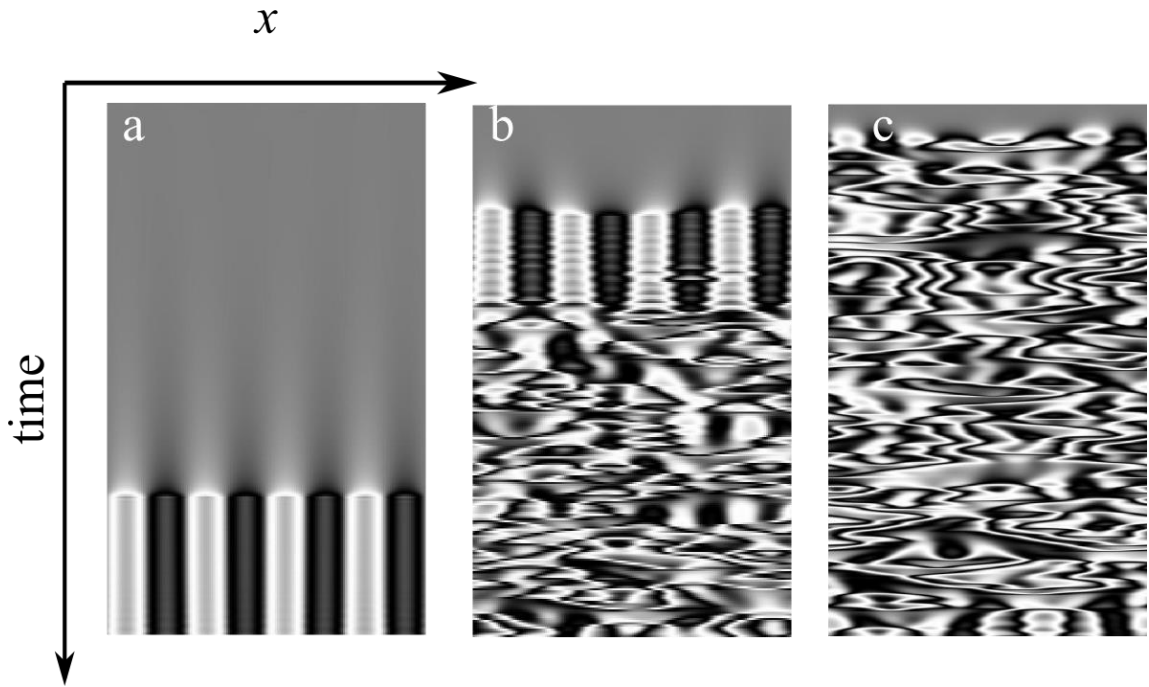

**Supplementary Figure 4. Space time diagrams obtained by numerical solution of Eqs. (1,2).** Grey-scale images show values of $\zeta = \sin(2(\phi - \theta))$, white color corresponds to $\zeta = 1$ and black to $\zeta = -1$. Parameters are $U = -0.3$, $K = 1$, $A = 1$, $\mu = 10$, $\eta = 1$, and for three values of the surface anchoring $\varepsilon = 0.05$ (a), 0.04 (b) and 0 (c). Domain of integration is 100 units, and time of integration 2000 units of time. Integration is performed in periodic boundary conditions; initial condition was small noise around $\phi = \theta = 0$ state.

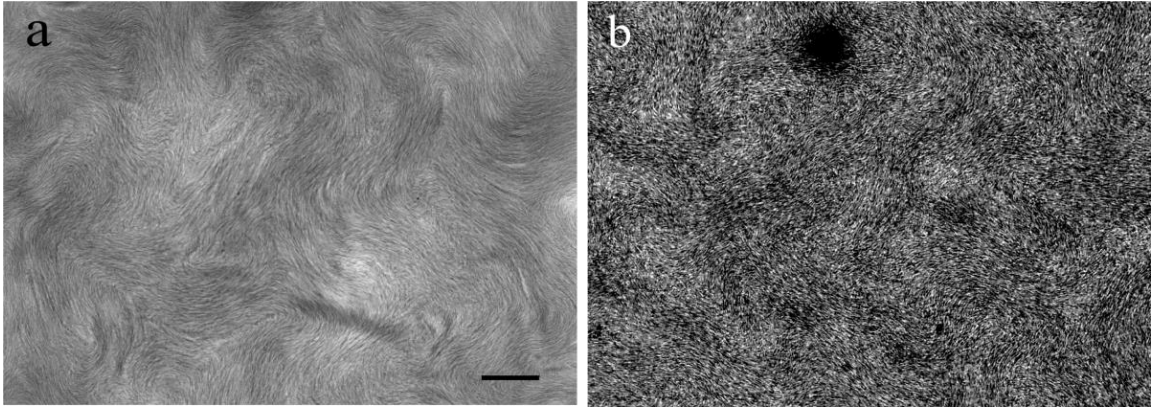

**Supplementary Figure 5. Comparison of collective motion in the LLC (a) and in the Newtonian liquid (b).** Image (a) displays collective motion in the LLC, pattern of bacterial and nematic director orientation shows multiple ±½ defect pairs (the same as Fig 3). Local almost perfect parallel orientation of bacteria is manifested by fine stripe texture. Image (b) shows snapshot of collective motion in water for similar concentration of bacteria. Unlike the texture (a), no fine stripe texture corresponding to local parallel alignment of bacteria can be identified in the part (b). Scale bar 30 μm for both images.